\documentclass[a4paper,10pt]{article}

\usepackage[dvips]{graphicx}
\usepackage{epsfig}

\def\jpb{{\em J. Phys. B: At. Mol. Opt. Phys.}~}
\def\pra{{\em Phys. Rev. A}~}
\def\prl{{\em Phys. Rev. Lett.}~}
\def\jmo{{\em J. Mod. Opt.}~}
\def\jetp{{\em Sov. Phys. JETP}~}
\def\etal{{\em et al.}}

\newcommand{\beq}{\begin{equation}}
\newcommand{\eeq}{\end{equation}}
\newcommand{\vecp}{{\bf p}}
\newcommand{\vecv}{{\bf v}}
\newcommand{\vecA}{{\bf A}}

\newcommand{\vecr}{{\bf r}}
\newcommand{\vecR}{{\bf R}}

\newcommand{\vecV}{{\bf V}}
\newcommand{\vecX}{{\bf X}}
\newcommand{\vecG}{{\bf G}}
\newcommand{\vece}{{\bf e}}

\begin{document}




\title{Strong Field Approximation for Systems with Coulomb Interaction}

\author{S.V. Popruzhenko$^{a,b}$ $^{\ast}$\thanks{$^\ast$ Corresponding author. Email: poprz@theor.mephi.ru\vspace{6pt}}
and D. Bauer$^{a}$\\\vspace{6pt} $^{a}${\em{Max-Planck Institut f\"{u}r Kernphysik, Heidelberg, Germany}};\\
$^{b}${\em{Moscow State Engineering Institut, Moscow, Russia}}}

\maketitle

\begin{abstract}
A theory describing above-threshold ionization of atoms and ions in a strong electromagnetic field is presented.
It is based on the widely known strong field approximation and incorporates the Coulomb interaction between the photoelectron and the nucleus using the method of complex classical trajectories.
A central result of the theory is the Coulomb-corrected ionization amplitude whose evaluation requires little extra numerical effort.
By comparing our predictions with the results of {\em ab initio} numerical solutions for two examples we show that the new theory provides a significant improvement of the Coulomb-free strong field approximation.
For the case of above-threshold ionization in elliptically polarized fields a comparison with available experimental data is also presented.
\bigskip


\end{abstract}

\section{Introduction}
Among analytical approaches developed for the description of strong field ionization of atoms, ions and molecules, the so-called strong field approximation (SFA) or Keldysh-Faisal-Reiss theory (KFR) \cite{keldysh,faisal,reiss} is known to be the most fruitful and widely applicable.
The objective of SFA is the description of above-threshold ionization (ATI) of atoms and ions in a strong electromagnetic field.
Being generalized to account for the effects of rescattering, SFA was used to calculate the high-energy plateau in ATI spectra, high-order harmonic generation and nonsequential multiple ionization.
The present state of the theory is reviewed in Refs.\cite{ago,becker02,popov04,faisal05}.
During the last years the scope of strong field physics has been extended to more complex systems including molecules, fullerenes and clusters where the SFA is also being used as a powerful tool (for SFA in molecules see, e.g., Ref.\cite{molecules} and references therein).

There are several formulations of the SFA which are not fully equivalent.
However, all of them are based on the idea that the electron continuum states can be approximated by plane Volkov waves.
The latter are exact solutions of the Schr\"odinger equation for a free electron in the field of a plane electromagnetic wave \cite{volkov}.
Using the plane Volkov continuum the matrix elements describing ionization and other related processes admit fast numerical evaluation.
Moreover, simple analytical treatments are possible in limiting cases.
Physically, the SFA fully disregards the effect of the binding potential on the continuum.
For negative ions (or, more general, for systems where electrons are bound by short-range forces), where the Coulomb interaction between the detached electron and the atomic core is absent, this approximation is well justified.
As a consequence, the SFA usually provides a quantitatively correct description of ATI and related processes in negative ions, as has been multiply proved by comparing SFA results with data \cite{kiyan01,kiyan05,becker07}, with predictions of a more rigorous quasistationary quasienergy states method \cite{manakov80,manakov-jpbr03,manakov-prl03,manakov-jpbl03} and with numerical solutions of the time-dependent Schr\"odinger equation (TDSE) for single active electron model atoms \cite{bauer05}.

In case of atoms, positive ions and molecules  the neglect of the long-range Coulomb interaction between the emitted electron and the ion core is not justified.
On a qualitative level the SFA often works for such systems too.
Several SFA calculations also led to good quantitative agreement with experimental data \cite{reiss96}.
However, it is now well established that the neglect of the Coulomb interaction may lead to essential contradictions.
Violation of the gauge invariance and wrong predictions for the static-field limit in the total (energy-angular integrated) ionization rate are widely known examples.
Besides, there are several well-documented features in ATI spectra which certainly cannot be explained within the standard SFA approach (see \cite{bashk,rudenko} for examples).

In recent years new opportunities to clarify the quality of different theoretical approaches are possible by {\em ab initio} numerical solutions of the time-dependent Schr\"odinger equation for one and two-electron systems driven by strong laser fields.
Compared to real experiments these {\it numerical experiments} provide more stringent tests for analytical theories since all parameters are exactly known and no focal averaging washes out interference effects which are sensitive to the details of the laser pulse shape and the carrier-envelope phase \cite{milos06}.
In general, the results of numerical experiments show that the SFA is only accurate on a qualitative level in the  description of ATI angular-resolved spectra, even in the simplest case of atomic hydrogen.
Many important details, including the interference structure, symmetry properties of photoelectron momentum distributions and the positions of ATI maxima reveal a disagreement between the SFA and TDSE results \cite{bauer05,bauer-jmo06}.
The key physical effect behind these disagreements is the long-range Coulomb interaction between the emitted electron and the atomic core.

Since the early days of laser physics considerable effort has been invested into taking the Coulomb interaction in strong field theories into account.
The first attempt was probably made in 1967 by Nikishov and Ritus \cite{nikishov67}.
In the same year, a Coulomb correction to the tunnel ionization rate of atoms was derived by Perelomov and Popov \cite{popov67}.
In the 80s the Coulomb-induced orders-in-magnitude enhancement of tunnel ionization rates of atoms and positive ions was well documented in experiments \cite{chin}.
Later, different models were proposed to account for the effect of the Coulomb force on the photoelectron during its propagation from the atom to a detector within classical mechanics.
Such calculations explained to some extend the Coulomb-induced asymmetries in elliptically \cite{gor04} and linearly \cite{chelk05} polarized fields; in the latter case an asymmetry may only appear in very short pulses \cite{milos06}.
Classical modeling of photoelectron trajectories in the presence of both the laser and the Coulomb fields has shown peculiarities in the extremely-low-energy part of ATI spectra \cite{chen,dimitriou}.
These peculiarities in the form of cusps and dips in photoelectron momentum distributions were recently observed in high-resolution experiments \cite{mosh03,rudenko}.
Within quantum approaches, so-called Coulomb-Volkov functions have been used for the description of Coulomb effects in ATI \cite{ferrante,ehlotzky,jaron,rudenko}.
For molecules, these wave functions were generalized in \cite{lein07}.
In Ref.\cite{faisal05l} the Coulomb-phase-distorted wave function was proposed and applied to the analysis of momentum distributions recorded in \cite{mosh03,rudenko}.
Finally, in Refs.\cite{smirnova06,smirnova08} a new eikonal-like approximation for the Coulomb-Volkov continuum was formulated and applied to the description of the subcycle ionization dynamics and the strong-field assisted XUV ionization.
However,  to the best of our knowledge up to now no reasonably simple theory of strong field ionization incorporating the Coulomb field in all its essential features has been put forward.

In this paper, we show how the SFA should be modified to incorporate Coulomb effects.
We call this new approach the  Coulomb-corrected strong field approximation (CCSFA).
The basic idea is to account for the Coulomb field by using semiclassical perturbation theory for the action.
In connection with the problem we consider this method was proposed in 1967 by Perelomov and Popov \cite{popov67}.
Here we generalize this idea for the description of angular-resolved ATI spectra.
This generalization exploits the imaginary time method \cite{popov05} or, equivalently, the technique of complex trajectories \cite{science,kopold02a,milos-jmo06}.
Applications of the proposed method demonstrate significant improvements of the theory. In fact, its predictions appear to be almost identical to the corresponding results of the {\em ab initio} single electron TDSE solution and yields  quantitative agreement with experimental data.
Although the theory requires a few more computations than the standard SFA, it remains  remarkably simple.

The paper is organized as follows.
The next section is devoted to the standard SFA ionization amplitude and to the semiclassical picture of complex classical trajectories.
In the third section a general method to include  Coulomb corrections is described and the CCSFA ionization amplitude is formulated.
The subsequent section shows two applications of the new theory.
The last section contains final remarks and conclusions. Atomic units are used unless noted otherwise.

\section{Coulomb-free Strong Field Approximation}
\subsection{SFA matrix element}
Within the SFA the transition amplitude between an atomic bound state $|\Psi_0\rangle$ of binding energy $\epsilon_0\equiv-I$ and a continuum state $|\Psi_{\vecp}\rangle$ with asymptotic momentum $\vecp$ is given by
\beq
M_{\rm SFA}(\vecp)=-i\int\limits_{-\infty}^{+\infty}\langle\Psi_{\vecp}\vert\hat V(t)\vert\Psi_0\rangle dt
\label{MSFA}
\eeq
where the final state is replaced by the Volkov function
\beq
\Psi_{\vecp}(\vecr,t)=\exp\left\{i\vecp\vecr-\frac{i}{2}\int\limits_{-\infty}^t(\vecp+\vecA(t_1))^2dt_1\right\}
\label{Volkov}
\eeq
and $\hat V(t)$ is the interaction operator which may be chosen in different gauges.
The laser field is described by the vector potential $\vecA(t)$ which depends only on time in the dipole approximation we use here.
The electric field is given by ${\cal E}(t)=-\partial_t\vecA$.

For a ground state in a short-range potential the amplitude (\ref{MSFA}) is gauge-independent, so that we equally may use the Volkov functions and the interaction operators in the length or velocity gauge.
If one starts from an {\em atomic} bound state $M_{\rm SFA}(\vecp)$ in general is  gauge-dependent, with sometimes very significant differences between length and velocity gauge.
For atoms the length gauge was shown to be the better choice \cite{bauer05,becker07} while for spatially extended systems such as molecules the velocity gauge works better \cite{molecules}.
However, in our following considerations we shall not be confronted with this unphysical gauge-dependence.
We start from the matrix element (\ref{MSFA}) for a state in a short-range well of ionization potential $I$ equal to the one of the real atomic state we consider.

Two dimensionless parameters (defined differently in the literature) turn out to be important within the SFA. 
Here we use the Keldysh parameter $\gamma$ and the multiquantum parameter $K_0$, defined as
\beq
\gamma=\kappa\omega/{\cal E}_0,~~~~~K_0=I/\omega
\label{gamma}
\eeq
where ${\cal E}_0$ and $\omega$ are the field amplitude and the laser frequency, respectively, and $\kappa=\sqrt{2I}$ is the characteristic momentum of the bound state.

The parameter $K_0$ defines the minimum number of photons necessary for ionization.
If $K_0\gg 1$ then, at arbitrary $\gamma$, the integral over time in (\ref{MSFA}) may be evaluated by the saddle-point method so that the amplitude can be represented as a sum of contributions from all relevant stationary points $t_s(\vecp)$,
\beq
M_{\rm SFA}(\vecp)=\sum_{\alpha}{\cal P}(\vecp,t_{s\alpha})\frac{\exp\left(-iS_0(\vecp,t_{s\alpha})\right)}
{\sqrt{\partial^2_tS_0(\vecp,t_{s\alpha})}},
\label{MSFAsp}
\eeq
where
\beq
S_0(\vecp,t)=\int\limits_t^{+\infty}\left\{\frac{1}{2}(\vecp+\vecA(t_1))^2+I\right\}dt_1
\label{S0}
\eeq
and the pre-exponential ${\cal P}$ contains the spatial matrix element.
The saddle-point equation has the form
\beq
\partial_tS_0(\vecp,t_{s\alpha})=\frac{1}{2}(\vecp+\vecA(t_{s\alpha}))^2+I=0
\label{speq}
\eeq
which shows that a saddle point $t_s(\vecp)$ is always complex for $I>0$.
The differential ionization rate is given by the square modulus of (\ref{MSFAsp}).

\subsection{Formulation in terms of complex trajectories}
The amplitude (\ref{MSFAsp}) can be equivalently reformulated in terms of classical complex trajectories \cite{popov05}.
For each stationary point consider a trajectory which satisfies the Newton equation in the laser field (below we omit a subscript $\alpha$ unless it is misleading)
\beq
\ddot\vecr_0=\dot\vecv_0=-{\cal E}(t)
\label{NE0}
\eeq
with the initial and boundary conditions
\beq
\vecv_0^2(t=t_s)=-\kappa^2,~~~~~\vecv_0(t\to +\infty)=\vecp.
\label{speq}
\eeq
The solution is
\beq
\vecr_0(\vecp,t)=\vecX_0+\vecp(t-t_s)+\vecG(t)-\vecG(t_s),~~~
\vecG(t)=\int\limits^t\vecA(t_1)dt_1.
\label{r0}
\eeq
Although a trajectory satisfies the classical equation of motion it is in general complex because of the initial condition.
However, the velocity $\vecv_0=\vecp+\vecA(t)$ is always real in real time.
We assume that the real part of a trajectory starts from the position of the atom, ${\rm Re}[\vecr_0(\vecp,t_s)]=0$.
Then, by a proper choice of a purely imaginary constant $\vecX_0$, we can make the coordinate real in real time too.
The trajectory (\ref{r0}) starts from $\vecX_0$ at $t=t_s$, having a purely imaginary velocity, $\vecv_0^2(\vecp,t_s)=-\kappa^2$.
As time approaches the real axis, $t=t_0\equiv{\rm Re}[t_s]$, both velocity and coordinate become real, so that the electron appears in real space at
\beq
\vecR_0(\vecp)=\vecr_0(\vecp,t_0)=\vecG(t_0)-{\rm Re}[\vecG(t_s)].
\label{R0}
\eeq
The position $\vecR_0$ is commonly interpreted as the ``tunnel exit''.
Although the tunneling picture of ionization is only relevant for $\gamma\ll 1$, the trajectory-approach is valid at arbitrary values of the Keldysh parameter.
Hence, we will use the term ``exit'' for (\ref{R0}) for arbitrary $\gamma$.
Beyond the exit, the electron propagates in real space and time towards a detector.
The values $\vecR_0$ and $\vecV_0\equiv\vecv_0(\vecp,t_0)$ should be considered as initial conditions for the motion in real time.
They fully determine the real part of a trajectory.

After the following algebra,
$$
S_0(\vecp)=\int\limits_{t_s(\vecp)}^{+\infty}\left(\frac{1}{2}\vecv_0^2+I+\frac{d}{dt}(\vecv_0\cdot\vecr_0)-\frac{d}{dt}(\vecv_0\cdot\vecr_0)\right)dt=
\nonumber
$$
$$
=-\int\limits_{t_s(\vecp)}^{+\infty}\left(\frac{1}{2}\vecv_0^2-\dot\vecv_0\cdot\vecr_0-I\right)dt+(\vecv_0\cdot\vecr_0)\vert_{t\to +\infty}-(\vecv_0\cdot\vecr_0)\vert_{t=t_s},
\nonumber
$$
we represent (\ref{MSFAsp}) in the form
\beq
M_{\rm SFA}(\vecp)=\sum_{\alpha}{\cal P}_{\alpha}(\vecp)\frac{\exp\left(iW_{0\alpha}(\vecp)\right)}
{\sqrt{\partial^2_tW_{0\alpha}(\vecp)}}
\label{MSFAW}
\eeq
where
\beq
W_0(\vecp)=\int\limits_{t_s(\vecp)}^{+\infty}({\cal L}_0-I)dt-\vecp\cdot\vecr_0(+\infty)+\vecv_0(t_s)\cdot\vecX_0
\label{W0}
\eeq
is the classical reduced action evaluated along the complex trajectory (\ref{r0}).
The corresponding Lagrangian is ${\cal L}_0=\vecv_0^2/2-{\cal E}(t)\cdot\vecr_0$.

The result (\ref{MSFAW}) shows that the SFA ionization amplitude can be represented as a coherent sum of contributions from complex classical trajectories.
To calculate each contribution we have to find a trajectory satisfying Eqs.(\ref{NE0}), (\ref{speq}) and evaluate the respective classical action (\ref{W0}).

\section{Coulomb-corrected Strong Field Approximation}

\subsection{Basic idea}
The above formulated representation of the SFA provides a natural way to include the Coulomb field via corrections to the classical action in (\ref{MSFAW}) evaluated along classical complex trajectories.
Assuming these corrections to be small compared to the laser-induced action (\ref{W0}) we may perform this procedure perturbatively taking Coulomb-free trajectories as a zero-order approximation \cite{popov67}.
There are two corrections. One is due to the appearance of the potential energy $U_{\rm C}$ in the action 
\beq
W_{\rm C}^{\rm (I)}(\vecp)=-\int\limits_{t_s(\vecp)}^{+\infty}U_{\rm C}[\vecr_0(\vecp,t)]dt={\cal Z}\int\limits_{t_s(\vecp)}^{+\infty}\frac{dt}{|\vecr_0(\vecp,t)|},
\label{SC1}
\eeq
another one is due to the correction $\vecr_1$ to a trajectory
\beq
W_{\rm C}^{\rm (II)}(\vecp)=\int\limits_{t_s(\vecp)}^{+\infty}[\vecv_0\cdot\vecv_1-{\cal E}(t)\cdot\vecr_1]dt.
\label{SC2}
\eeq
Here the correction $\vecr_1(\vecp,t)$ has to be determined from the Newton equation
\beq
\ddot\vecr_1=-\frac{{\cal Z}(\vecr_0+\vecr_1)}{|\vecr_0+\vecr_1|^3}
\label{NE}
\eeq
with ${\cal Z}$ being the residual charge (${\cal Z}=1$ for a neutral atom).
Besides, the stationary point $t_s(\vecp)$ may also be corrected.

This general idea meets, however, with several difficulties:\\
(i) Corrections (\ref{SC1}) and (\ref{SC2}) are logarithmically divergent at the lower integration limit $t\to t_s$ when the electron approaches the nucleus because for the most important trajectories the imaginary starting point $\vecX_0$ in (\ref{r0}) is small or exactly equal to zero.
Obviously, under such circumstances the singular Coulomb force cannot be treated perturbatively.\\
(ii) A zero-order trajectory $\vecr_0(\vecp,t)$ may revisit the nucleus in real time (such trajectories are known to be responsible for the rescattering phenomena).
This causes another divergence.\\
(iii) It is unclear which initial conditions should be attached to Eq.(\ref{NE}) for the correction $\vecr_1(\vecp,t)$.

The problem (i) was solved already in Ref.\cite{popov67}.
Below we show how to resolve the other two by the proper choice of a zero-order trajectory.

\subsection{A choice of the zero-order trajectory}
If we take a trajectory $\vecr_0(\vecp,t)$ and evaluate a correction in real time by solving (\ref{NE}) for $t\ge t_0$ then, even if the electron does not revisit the origin and the correction is everywhere small, we obtain a different final photoelectron momentum $\vecv(t\to+\infty)\ne\vecp$.
Since, however, we want to find the ionization amplitude for a given final momentum $\vecp$, the above described procedure does not appear satisfactory.

A correct realization is as follows.
Assume that we know some other Coulomb-free trajectory $\vecr_0(\tilde{\vecp},t)$ such that, if we account for the Coulomb force at $t\ge\tilde{t}_0$, the final momentum will exactly have the desired value $\vecp$.
Below we denote such a trajectory $\vecr_0(\tilde{\vecp},t)\equiv\tilde{\vecr}_0(\vecp,t)$.
A corresponding stationary point $\tilde{t}_s({\vecp})$ satisfies Eq.(\ref{speq}) with $\tilde{\vecp}$ instead of $\vecp$ so that
\beq
\tilde{t}_s(\vecp)=t_s(\tilde{\vecp}).
\label{speqnew}
\eeq
The new initial conditions for the motion in real time have the form
\beq
\tilde{\vecR}_0(\vecp)=\vecR_0({\tilde{\vecp}})=\vecG(\tilde{t}_0)-{\rm Re}\vecG(\tilde{t}_s),~~~
\tilde{\vecV}_0(\vecp)=\tilde{\vecp}+\vecA({\tilde{t}_0}).
\label{icnew}
\eeq

How to find this trajectory?
First, we give some physically transparent arguments which explain why such a trajectory and the respective renormalized momentum $\tilde{\vecp}$ always exist.
We start the electron's propagation at $t=\tilde{t}_0$ when the electron is at the ``tunnel exit'' $\tilde{\vecR}_0$.
Here the Coulomb force is already small compared to the laser field  (for an exact criterion see Subsection 3.4).
Therefore, if the electron does not approach the nucleus, a Coulomb-induced deformation of the trajectory is smooth.
Thus, just according to physical continuity each trajectory $\vecr_0(\vecp,t)$ must have an analog $\tilde{\vecr}_0(\vecp,t)$ which, without the Coulomb force corresponds to some different final momentum $\tilde{\vecp}$ and goes to $\vecp$ if the Coulomb field is turned on.
Moreover, even if the initial trajectory is approaching the nucleus, the new one will not because of the following two reasons.\\
1. In full dimensionality the fraction of trajectories which closely encounter the nucleus is very small.
Even in a linearly polarized field the relative amount of trajectories which approach the parent ion at distances of several atomic units typically is $10^{-4}\div 10^{-6}$.
This value is nothing else but the relative probability of rescattering, known to be always small.\\
2. If a trajectory approaches the nucleus, it appears to be strongly disturbed by the Coulomb force and, as a result, the electron in general will contribute to the high-energy rescattering part of the spectrum we are not interested in here.
Therefore, such trajectories, although existing, cannot be those we are looking for.

In a more strict formulation, although some fraction of trajectories appear to be strongly disturbed by the Coulomb field and show {\em irregular behavior}, for any given value of the final momentum $\vecp$ new trajectories $\tilde{\vecr}_0(\vecp,t)$ exist which, after the Coulomb field is taken into account, experience only a smooth perturbation.
The number of such {\em regular} trajectories is equal to the number of initial trajectories $\vecr_0(\vecp,t)$.

\begin{figure}
\includegraphics[width=13cm]{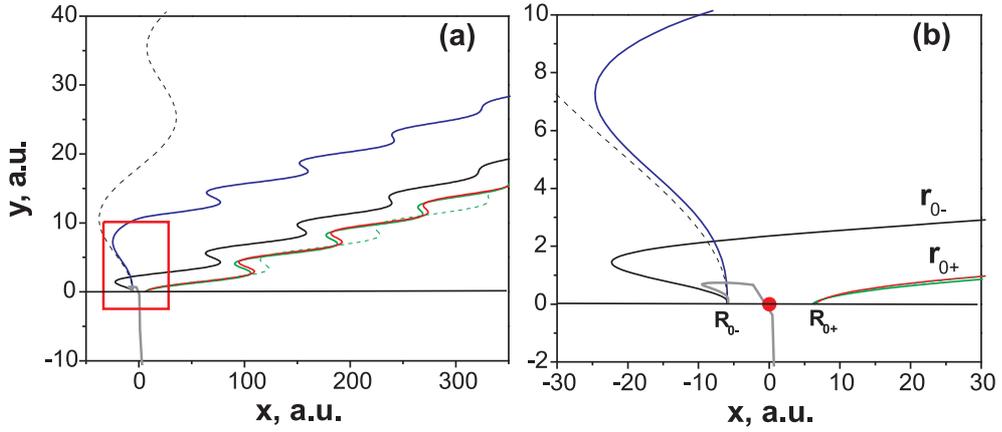}
\caption{(Color online) Panel (a): Trajectories corresponding to ionization from the ground state of hydrogen into the final state with energy $\epsilon=p^2/2=0.278$ (5-th ATI maximum) and emission angle $\theta=0.05$ in the field (\ref{Elin}) with ${\cal E}_0=0.08$ and $\omega=0.056$. The emission angle is chosen different from zero just to make the Coulomb-free trajectories $\vecr_{0\pm}(\vecp,t)$ two-dimensional and therefore visible in the plot. They  are shown by green and black solid lines, the respective trajectories $\tilde{\vecr}_{0\pm}(\vecp)$  by dashed lines. The full solutions of Eq.(\ref{NE}) $\vecr_{\pm}(\vecp,t)$ with initial conditions (\ref{icnew}) are shown by red and blue. An irregular trajectory developing from $\vecr_{0-}$ under the action of the Coulomb force is shown in gray. Panel (b) shows the space indicated by a red rectangle in panel (a) with more resolution. The position of the atom is indicated by a red circle. It is seen that the trajectory which does not return experiences only a small distortion while  the other trajectory is changed significantly. The values of the final and the renormalized momenta are $\vecp\equiv(p_x,p_y)=(0.745,0.037)$, $\tilde{\vecp}_+=(0.971,0.043)$, $\tilde{\vecp}_-=(0.369,0.223)$.}
\label{fig1}
\end{figure}
We illustrate this statement for the case of a linearly polarized field where the most important trajectories may approach the nucleus closely so that this example is the most challenging one.
Consider ionization in the field 
\beq
{\cal E}(t)={\cal E}_0\cos(\omega t),
\label{Elin}
\eeq
polarized along the $x$-axis.
For a constant field envelope all laser periods are identical.
For each final momentum $\vecp$ there are two trajectories per period (we denote them $\vecr_{0+}(\vecp,t)$ and $\vecr_{0-}(\vecp,t)$), whose motion in real time starts from the points $\vecR_{0+}$ and $\vecR_{0-}=-\vecR_{0+}$ symmetrically displaced with respect to the atom.
If the transverse momentum $p_y$ is small or equal to zero the trajectory $\vecr_{0+}$ will never return while $\vecr_{0-}$ revisits the nucleus at least once (see Fig.1).
It is intuitively clear that the Coulomb force along the first trajectory is always small and simply decelerates the electron on its way to a detector.
Therefore, the new Coulomb-free trajectory $\tilde{\vecr}_{0+}$ will be quite close to the initial one and the renormalized momentum $\tilde{\vecp}_+$ is expected to be larger in absolute value than $\vecp$ in order to compensate this deceleration.
In contrary, the electron moving along $\vecr_{0-}$ will accumulate a significant negative transverse momentum due to the Coulomb attraction when it passes near the nucleus.
Using $\vecr_{0-}$ as a zero approximation we obtain a typical irregular trajectory shown in Fig.1 in  gray.
Thus, to arrive at a detector with a prescribed small transverse momentum, the electron has to start with some significant one, $\tilde{p}_y>0$.
However this immediately leads to the trajectory $\tilde{\vecr}_{0-}$ which stays away from the nucleus and remains  regular.
This example, summarized in Fig.1, shows that even in a linearly polarized field the regular Coulomb corrected trajectories exist and they can be obtained from initial Coulomb-free trajectories by a smooth deformation.
The key point is to find proper zero-order trajectories $\tilde{\vecr}_0$ instead of $\vecr_0$.
In an elliptically polarized field, where trajectories are always two-dimensional, the situation is even simpler.

Another question is how to find these trajectories in a particular calculation.
Except limiting cases this can only be done by solving Eq.(\ref{NE}) numerically.
Since not the initial but the final condition $\vecv(t\to+\infty)=\vecp$ is fixed we have to deal with an {\em inverse problem}.
Because of this additional complication a proper trajectory can only be found, to the best of our knowledge, by iterations starting from some appropriate zero-order approximation.
However, this numerical iteration is simple to implement and the
evaluation of one Coulomb-corrected trajectory takes a fraction of a second on a PC using a standard solver for differential equations.
In the two applications shown below two different iteration procedures have been applied, both leading rapidly to converged results.

\subsection{Sub-barrier corrections and the matching procedure}
The algorithm described in the previous subsection is incomplete since corrections during the ``sub-barrier'' motion while  $t\in [t_s,t_0]$ should also be taken into account.
Because a trajectory $\tilde{\vecr}_0$ already yields, after accounting for the Coulomb force at $t\ge t_s$, the desired final momentum $\vecp$, the boundary conditions for sub-barrier motion are $\vecr_1(\tilde{t}_0)=0,~\vecv_1(\tilde{t}_0)=0$.
Consider the most ``dangerous'' case when $\vecX_0=0$ and all the values $\vecr_1$, $\vecv_1$, $W_{\rm C}^{\rm (I)}$ and $W_{\rm C}^{\rm (II)}$ are divergent at $t\to t_s$.
The regularization procedure is based on the idea that, as the electron approaches the nucleus, its dynamics should be determined mainly by the binding force, not by the laser field \cite{popov67}.
Thus, the respective action $W_{\rm at}$ can be found from an asymptotic expression for the atomic bound state wave function at large distances $r\gg 1/\kappa$.
For $s$-states this asymptotic behavior is
$$	
\Psi_0(r)\sim(\kappa r)^{n_*}\exp\{-\kappa r\}=\exp\{-\kappa r+n_*\ln(\kappa r)\}\equiv\exp\{iW_{\rm at}(r)\},
\nonumber
$$
so that
\beq
W_{\rm at}(r)=i[\kappa r-n_*\ln(\kappa r)].
\label{Sat}
\eeq
Here, $n_*={\cal Z}/\kappa$ is the effective principle quantum number of the bound state \cite{popov04}.

To match the Coulomb-corrected action with (\ref{Sat}) we extract the divergent part of the former at $t\to t_s$.
To this end we introduce a matching point $t^*\in [t_s,t_0]$ such that $1/\kappa\ll|\vecr_0(\vecp,t^*)+\vecr_1(\vecp,t^*)\equiv\vecr^*|\ll R_0(\vecp)$.
Here, the laser-induced action (\ref{W0}) has the behavior
\beq
W_0(t^*)\approx W_0(\vecp)-2I(t_s-t^*).
\label{ass1}
\eeq
Integrating (\ref{SC2}) by parts and taking into account (\ref{NE0}) and the initial conditions $\vecr_1(\tilde{t}_0)=\vecv_1(\tilde{t}_0)=0$ we obtain
\beq
W_{\rm C}^{\rm (II)}(t^*)=-i\kappa(\vece(t_s)\cdot\vecr_1^*),~~~~~\vece(t_s)\equiv{\cal E}(t_s)/|{\cal E}(t_s)|.
\label{ass2}
\eeq
Finally, the divergent part of (\ref{SC1}) is
\beq
W_{\rm C}^{\rm (I)}(t^*)\approx -in_*\ln\frac{2i\gamma}{t_s-t_*}.
\label{ass3}
\eeq
At $t^*\to t_s$ the solution of Eq.(\ref{NE0}) with the initial conditions $\vecr_0(t_s)=0,~\vecv_0^2(t_s)=-\kappa^2$ is
\beq
\vecr_0^*\equiv\vecr_0(t^*)\approx -i\kappa\vece(t_s)(t_s-t^*).
\label{x0*}
\eeq
Now, collecting the expressions (\ref{Sat}), (\ref{ass1}), (\ref{ass1}), and (\ref{ass2}) and (\ref{x0*}),
we see that indeed all terms depending on the matching point disappear in the full action $W_{\rm at}(r^*)+W_0(t^*)+W_{\rm C}^{\rm (I)}(t^*)+W_{\rm C}^{\rm (II)}(t^*)$ and the final result is convergent.

The regular part of (\ref{SC2}) is equal to zero, the one of (\ref{SC1}) can be evaluated for arbitrary field polarization and photoelectron final momentum only numerically.
In a linearly polarized field, for the most probable trajectory with $\vecp=0$, it is \cite{popov67}
$$
{\rm Reg}[W_{\rm C1}](\vecp=0)=-in_*\ln(2\kappa^3/{\cal E}_0)\equiv-in_*\ln(2/F),
\nonumber
$$
where $F={\cal E}_0/\kappa^3$ is the reduced laser field; the value ${\cal E}_{\rm at}=\kappa^3$ has the meaning of a characteristic atomic field.
This contribution causes the well-known Coulomb correction enhancing the total ionization rate \cite{popov67}.
The regularization procedure described above becomes more cumbersome, although doable, for a laser field of arbitrary polarization and for an arbitrary final momentum if $\vecX_0(\vecp)\ne 0$.
Also bound states of arbitrary angular momentum can be treated.

\subsection{CCSFA ionization amplitude and applicability conditions}
Summarizing the results obtained in this section we formulate an algorithm for the evaluation of the CCSFA ionization amplitude:\\
1. For a given value of the final momentum $\vecp$ determine all relevant saddle points $t_{s\alpha}(\vecp)$ and the respective trajectories $\vecr_{0\alpha}(\vecp,t)$.\\
2. For each trajectory find the renormalized Coulomb-free trajectory $\tilde{\vecr}_{0\alpha}$, the stationary point $\tilde{t}_{s\alpha}$, the momentum $\tilde{\vecp}_{\alpha}$ and the Coulomb corrected trajectory $\vecr_{\alpha}$ which is a solution of the Newton equation in two fields with the initial conditions (\ref{icnew}) by solving the inverse problem using an iteration scheme.\\
3. Calculate the Coulomb-free action (\ref{W0}) for the new trajectory $\tilde{\vecr}_{0\alpha}$. Here one should note that the renormalized momentum $\tilde{\vecp}_{\alpha}$ enters only the imaginary part of the action since the latter is accumulated for $t\in[\tilde{t}_{s\alpha},\tilde{t}_{0\alpha}]$, while the real part of the action depends on the final momentum $\vecp$.\\
4. Calculate the regular part of the Coulomb-induced action (\ref{SC1}) along the exact trajectory $\vecr_{\alpha}$.\\
Finally, the CCSFA ionization amplitude has the form
\beq
M_{\rm CCSFA}(\vecp)=\sum_{\alpha}{\cal P}_{\alpha}(\vecp)\frac{\exp\left(i[\tilde{W}_{0\alpha}(\vecp)+W_{{\rm C}\alpha}^{\rm (I)}(\vecp)]\right)}
{\sqrt{\partial^2_tW_{0\alpha}(\vecp)}}
\label{CCMSFA}
\eeq
where
$$
{\rm Im}[\tilde{W}_{0}(\vecp)]\equiv{\rm Im}[W_0(\tilde{\vecp},\tilde{t}_s)],~~~~~
{\rm Re}[\tilde{W}_{0}(\vecp)]\equiv{\rm Re}[W_0(\vecp,\tilde{t}_s)].
\nonumber
$$
Since the pre-exponential is much less significant we neglect corrections to it in this work.

The amplitude (\ref{CCMSFA}) is relevant only if the formalism of trajectories is applicable.
This requires the validity of the saddle-point method for the evaluation of the Coulomb-free amplitude (\ref{MSFA}), i.e.
$K_0\gg 1$.
Moreover, the Coulomb force at the ``exit'' (\ref{R0}) must be small compared to the laser field amplitude.
Using the reduced field $F$ and the Keldysh parameter $\gamma$ the latter restriction can be formulated as
\beq
1\gg F\times
\left\{\begin{array}{l}\displaystyle 1,~~~~~~\gamma\ll 1\\
\displaystyle\gamma^2,~~~~~\gamma\gg 1\end{array}.
\right.
\label{mu}
\eeq
The condition (\ref{mu}) shows that the standard SFA can only be used as a zeroth-order approximation for the development of a quantitatively correct theory of strong field ionization in the case of relatively weak fields ${\cal E}_0\ll {\cal E}_\mathrm{at}$ of not too high frequency $\omega\ll I\sqrt{F}$.
In terms of the laser field amplitude, the condition (\ref{mu}) can be reformulated as $1/K_0^2\ll F\ll 1$.
The corresponding intensity domain does not coincide with the one determining the tunneling regime, $\gamma\ll 1$.
The upper restriction on the Keldysh parameter $\gamma\ll 1/\sqrt{F}$ is much softer and allows treatments in the multiphoton regime.
Of course, for $F\to 0$ or $\gamma\to\infty$ condition (\ref{mu}) will be violated, so that the amplitude (\ref{CCMSFA}) does not match the perturbation theory result.
This feature is known to be inherent in the standard SFA as well.

\section{Applications}
In this section we evaluate the CCSFA ionization amplitude (\ref{CCMSFA}) and calculate the respective spectra for elliptically and linearly polarized fields and compare our results with the predictions of the standard SFA and those obtained by numerically solving the time-dependent Schr\"odinger equation in three spatial dimensions for atomic hydrogen, neon and argon using the code described in \cite{bauer-koval}.
The TDSE is solved for an atom with a single active electron whose bound state is characterized by the binding energy, orbital and magnetic quantum numbers.
The self-consistent effective single electron potential $U(r)$ may be constructed using the density functional method \cite{dft} or approximating it by a screened Coulomb potential with parameters adjusted to reproduce both the correct asymptotic behavior $U(r\gg 1/\kappa)\sim U_{\rm C}(r)=-1/r$ and the ionization potential of the ground state.
The pulse with the carrier frequency $\omega$, amplitude ${\cal E}_0$, ellipticity $\xi$ and a duration of $n=8$ optical cycles polarized in the $(x,y)$-plane is described by the vector potential
\beq
\vecA(t)=\frac{{\cal E}_0}{\omega\sqrt{1+\xi^2}}\left[\sin^2\left(\frac{\varphi}{2n}\right)\cos\varphi,
-\xi\sin^2\left(\frac{\varphi^{\prime}}{2n}\right)\cos\varphi^{\prime}\right]
\label{A}
\eeq
with $\varphi\equiv\omega t$, $\varphi^{\prime}=\varphi-\pi/2$ and $0\le\varphi,\varphi^{\prime}\le 2\pi n$.

\subsection{Symmetry violation in an elliptically polarized field}
It is now well established that, while the SFA predicts a fourfold symmetry of the photoelectron distribution for an arbitrarily polarized laser field (if finite pulse duration-effects are neglected), the experiments show that in elliptically polarized laser light this symmetry is clearly broken \cite{bashk,paulus00}.
According to the intuitive semiclassical ionization picture \cite{corkum89}, which often is in  remarkable agreement with the predictions of the SFA, the most probable photoelectron momentum is along the minor polarization axis of the polarization ellipse for intermediate ellipticities, say, $0.2 < \xi < 0.8$.
This effect of {\em classical dodging} \cite{pptII,b98} simply follows from the fact that the electron tunnels out with  zero velocity at the ``tunnel exit''  with highest probability near the maximum of the electric field.
At  that time the vector potential is oriented along the minor polarization axis, determining the drift photoelectron momentum $\vecp=\vecv-\vecA$ measured at a detector.
The SFA angular distributions shown in Fig.2 by green lines clearly demonstrate this effect.
However, the experimental results do not display the fourfold symmetry at all: photoelectrons go predominantly along the {\em major} polarization axis and the distribution possesses inversion symmetry only.
A typical distribution recorded in \cite{paulus00} is shown in Fig.2(d).

\begin{figure}
\includegraphics[width=11cm]{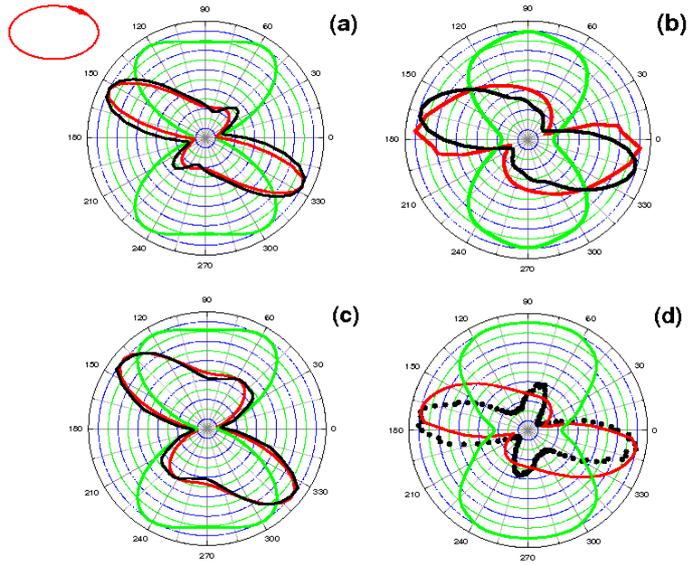}
\caption{(Color online) Normalized angular distributions in the polarization plane, evaluated by different methods: the standard SFA (green), CCSFA (red) and the {\em ab initio} TDSE solution (black) for ground states of hydrogen (a,b), neon (c) and argon (d). In the latter case the data recorded in \cite{paulus00} are shown by black circles while the TDSE result is not shown. The laser intensity, frequency, ellipticity and the photoelectron energy are: $1.0\cdot 10^{14}$W/cm$^2$, 1.55eV, 0.5 and 8.2eV (a), $1.6\cdot 10^{14}$W/cm$^2$, 3.1eV, 0.5 and 2.6eV (b), $2.0\cdot 10^{14}$W/cm$^2$, 1.55eV, 0.36 and 7.1eV (c) and $0.6\cdot 10^{14}$W/cm$^2$, 1.55eV, 0.36 and 2.8eV (d). For argon and neon the ionization probability is averaged over the magnetic quantum numbers $m=0,\pm 1$ within the p-shell.
The orientation of the polarization ellipse is shown in the insert with the rotation direction of the electric field vector indicated by an arrow.}
\label{fig2}
\end{figure}
Already in early works \cite{bashk,ferrante} it was suggested that the asymptotic Coulomb interaction of the outgoing photoelectron with its parent ion breaks the fourfold symmetry.
However, except \cite{gor04}, no explicit theory-{\em vs}-experiment analysis was presented.
Although the results obtained in Ref.\cite{gor04} demonstrate reasonable agreement with the data, the lack of {\em ab initio} calculations did not permit to draw definite conclusions.
Now we may present the full analysis along the line standard SFA {\em vs} CCSFA {\em vs ab initio} TDSE results {\em vs} experiment.
The iterative procedure in CCSFA was performed taking the trajectories $\vecr_0(\vecp,t)$ as a zero approximation and turning the nuclear charge on smoothly.
Figure~2 shows polar plots of the angular distributions for fixed ATI maxima calculated for different atoms and field parameters.
Panel (d) shows the data for argon recorded in Ref.\cite{paulus00}.
The standard SFA with its fourfold symmetry is in strong disagreement with both experimental data and numerical results.
Agreement between CCSFA and {\em ab initio} TDSE results is everywhere quantitatively good, particularly for the cases (a,c) where the parameter (\ref{mu}) is smaller.
The agreement with the data is also satisfactory, although neither CCSFA nor {\em ab initio} calculations (the latter is not shown in Fig.2(d) because it is almost identical to the CCSFA result) show the well-developed second maximum which is  clearly visible in the data.
Among other possible reasons for this disagreement we think it is due to the  uncertainty in the experimental laser intensity.
The intensity declared in Ref.\cite{paulus00} was $0.9\cdot 10^{14}$W/cm$^2$.
In calculations we used $0.6\cdot 10^{14}$W/cm$^2$ which led to better agreement including the  signature  of the second maximum (red curve).

\subsection{Interference in a linearly polarized field}
In a field of linear polarization, at least two trajectories per laser period contribute to the amplitude with comparable weights (exactly equal for the field (\ref{Elin}) having a constant envelope).
A number of relevant trajectories $> 1$ leads to the interference structure visible in ATI spectra and was first considered in \cite{pptI}.
With increasing field ellipticity, for the vast majority of final momenta, the contributions from the above mentioned trajectories become significantly different in absolute value, and interferences disappear.
In short pulses, the interference structure depends also upon the absolute phase, as it was observed in the attosecond double-slit experiment \cite{attoslit}.
{\em Ab initio} TDSE calculations performed in Ref.\cite{attoslit} showed good quantitative agreement with the data.
The SFA results, however, reproduced the interference structure only on a qualitative level.
An example shown in Fig.3 demonstrates the obvious quantitative disagreement between TDSE and SFA: interference maxima and minima in SFA and TDSE results are almost opposite to each other, meaning that the interfering trajectories in plain SFA have a relative phase which is wrong by $\pi$.

\begin{figure}
\includegraphics[width=11cm]{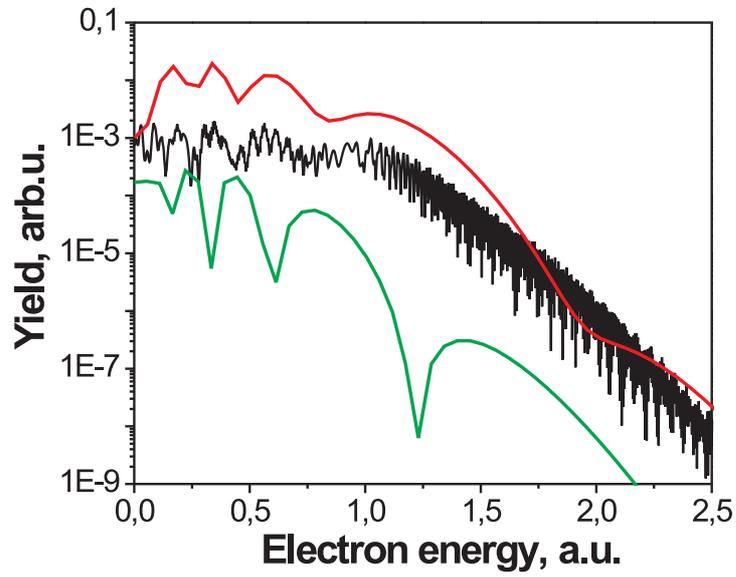}
\caption{(Color online) Photoelectron spectrum along the polarization direction calculated from the standard SFA (green), CCSFA (red) and the {\em ab initio} TDSE solution (black) for ionization from the ground state of hydrogen by a linearly polarized field with ${\cal E}_0=0.084$ ($2.5\cdot 10^{14}$W/cm$^2$) and $\omega=0.056$ (1.55eV).
The TDSE solution is performed for an 8-cycle pulse (\ref{A}) while the SFA and CCSFA amplitudes were calculated for the field (\ref{Elin}) of constant envelope.
The spectra are shifted vertically for visual convenience.}
\label{fig3}
\end{figure}
The spectrum calculated using the CCSFA (red line in Fig.3) demonstrates good quantitative agreement with the TDSE result, except for very low energies.
This shows that the shift of the interference maxima and minima should be entirely attributed to the effect of the Coulomb field.
As we found in calculations, the most significant contribution to the shift is coming from the real part of (\ref{SC1}) induced by the Coulomb interaction in the Lagrangian.
Since the action $W_{\rm C}^{\rm (I)}$ is logarithmically divergent at the upper integration limit one should calculate it along an exact trajectory $\vecr_{\pm}(\vecp,t)$ where the asymptotic momentum is $\vecp$.
Only in this case the unphysical divergent parts cancel each other in the phase difference of the interfering trajectories.
For the calculation of the spectrum along the polarization direction, as it was done for Fig.3, the trajectories $\vecr_0(\vecp,t)$ cannot be taken as a zero approximation in the iteration procedure since half of them go exactly through the origin.
According to Subsection 3.2, the proper trajectories $\tilde{\vecr}_0$ should carry a nonzero transverse momentum.
Therefore, trajectories which enter the iteration procedure must also have some nonzero transverse momentum.
In this case the iteration procedure does not require a smooth turn-on of the nuclear charge (as it was done for elliptical polarization).
The procedure yields rapidly converged trajectories, and
the renormalized transverse momentum $\tilde{p}_y$ does not depend on the particular (non-vanishing) seed value for the transverse momentum.

\subsection{Other possible applications}
There are at least two other straightforward applications of the theory.\\
1. With the development of intense sources of coherent XUV radiation based on free electron lasers delivering photon energies $\hbar\omega\simeq 10$eV and higher \cite{FEL}, the experimental study of strong field ionization of atoms and positive ions as well as other related processes in the intermediate $(\gamma\simeq 1)$ and multiphoton $\gamma\gg1 $ regimes became possible \cite{wabnitz02,laarmann04,wabnitz05}.
In particular, ionic charge states up to Xe$^{21+}$ were detected from a xenon gaseous target subject to 13.4\,nm wavelength pulses of intensities about $10^{16}$W/cm$^2$ \cite{sorokin}.
Under such conditions a typical value of the Keldysh parameter is $\gamma=10\div 50$.
Even the single electron ionization dynamics is still poorly understood in this frequency domain.
The method introduced in this paper can be applied to the evaluation of the Coulomb correction to the total ionization rate in the multiphoton regime, $\gamma\gg 1$, i.e., for the above described conditions.
This task was partially considered in \cite{popov07} where it was shown that this correction may be even larger in absolute value than in the tunneling limit (where it is also substantial \cite{popov67}).\\
2. The method can also be used to calculate the probability of excitation into a Rydberg state, instead of ionization.
Indeed, in short pulses the electron liberated from the atom and propagating towards a detector may have a negative total energy when the pulse is off.
In this case such an electron will be captured in  Rydberg states.
The effect of anomalously high Rydberg population was seen in {\em ab initio} simulations and recently found in an experiment \cite{becker08}.
In order to obtain a distribution over the Rydberg levels one should apply the algorithm described above assuming a given negative energy and an angular momentum corresponding to a Rydberg state, instead of a final photoelectron momentum $\vecp$ corresponding to ionization.

\section{Conclusions}
The Coulomb corrections we introduced here are based on trajectories and are therefore gauge invariant.
As a result the CCSFA amplitude is entirely gauge invariant for ionization from a  $s$-ground state.
For excited states, there is some gauge dependence appearing via the angular part of the bound state wave function $\Psi_0(\vecr)$ in a short-range well we use to calculate the pre-exponential in (\ref{CCMSFA}).
This dependence is, however, much weaker than the one which appears in the standard SFA if an atomic wave function is being used in (\ref{MSFA}).

The application of our method to the strong field ionization of molecules seems to be the most natural generalization of the theory.
Besides, per construction, CCSFA is applicable to any system where a long-range single particle or self-consistent force is important.
Among such systems small metal clusters and fullerenes can be mentioned.
Moreover, since in these systems the Coulomb interaction is not singular, evaluation of the CCSFA amplitude should be an even simpler task than it is for atoms and molecules.
Finally, the method can be applied not only to ionization but also to other strong field processes involving charged particles.

\section{Acknowledgment}
This work would never have appeared without intensive debates on the Coulomb problem in ionization the authors had with W.~Becker, S.P.~Goreslavski, V.D.~Mur, V.S.~Popov and H.R.~Reiss over the past decade.
We also thank A.~Bandrauk, C.~Chiril\v{a}, M.V.~Frolov, M.~Ivanov, N.L.~Manakov, D.B.~Milo\v{s}evi\'c, N.I.~Shvetsov-Shilovski, O.~Smirnova, A.F.~Starace, and D.F.~Zaretsky for fruitful discussions and interest.
We are grateful to G.G.~Paulus for providing us the data recorded in \cite{paulus00}.

The work was supported by the Deutsche Forschungsgemeinschaft.


\end{document}